\documentclass[aps,prd,superscriptaddress,twocolumn]{revtex4-1}

\usepackage{graphicx}
\usepackage{amsmath}
\usepackage{amssymb}
\usepackage{color}
\usepackage[per-mode = fraction]{siunitx}
\usepackage{chemformula}
\usepackage{bm}

\begin{document}
	
\title{Nernst and Ettingshausen effects in the Laughlin geometry}
\author{S.G.~Sharapov}
\affiliation{Bogolyubov Institute for Theoretical Physics, National Academy of Science of Ukraine, 14-b Metrologichna Street, Kyiv, 03143, Ukraine}
\author{A.A.~Varlamov}
\affiliation{CNR-SPIN, c/o DICII-University of Rome ``Tor Vergata'', Via del Politecnico, 1, 00133 Rome, Italy}
\author{C.~Goupil}
\affiliation{Universit\'{e} de Paris, LIED, CNRS UMR 8236, 5 Rue Thomas Mann, 75013 Paris, France}
\author{A.V.~Kavokin}
\affiliation{Westlake University, 18 Shilongshan Road, Hangzhou 310024, Zhejiang Province, China}
\affiliation{Westlake Institute for Advanced Study, Institute of Natural Sciences, 18 Shilongshan Road, Hangzhou 310024, Zhejiang Province, China}

\begin{abstract}
The ideal reversible thermodynamic cycle visualization of the Nernst effect in Laughlin geometry,
excluding  the kinetic contribution is proposed. The Ettingshausen effect is also treated in the fashion using the reverse cycle.
The corresponding values of the off-diagonal thermoelectric coefficients
are expressed  through the ratio of the entropy budget per magnetic flux.
Our approach enlightens the profound thermodynamic origin of the relation between the Nernst effect and magnetization currents.
\end{abstract}

\date{ \today}

\maketitle

\section{Introduction}

Two centuries ago Sadi Carnot introduced the notion of the ideal heat engine with molecular gas as the working body \cite{Carnot1824}.
The molecular gas changes its state  performing the closed cycle, which consists of two adiabatic and two isothermic curves in the restricting
area between the four points in the pressure $P$ and volume $V$ plane.

At the same time Johann Seebeck discovered the appearance of the  potential difference across a hot and cold end
for two dissimilar metals, which allowed  the creation of the thermoelectric generators afterward.
It was demonstrated  much later that the operation of the thermoelectric couple can be described
(see, e.g., the textbooks in Refs.~\cite{Samoylovich2007book,Goupil2015book}) in complete analogy with
the Carnot heat engine: it is enough to replace the molecular gas
by the degenerated Fermi gas as the working body and notice that the role of pressure in this case plays the electrochemical potential
\cite{Apertet2016EPJPlus}
$\tilde{\mu}=\mu+e \varphi$ (here $\mu$ is chemical potential and $e=-|e|$ is electron charge),
while instead  volume one implies the number of particles $N$.
Accordingly, the role of work $- P dV$ is played by
the energy of the mass transfer part of the first law of thermodynamics $\mu dN$
(the ``mass action'', as formulated in the classical textbook of Ref.~\cite{Kubo1968book}, see also Ref.~\cite{Vining1997}).
Here and in the following we set $\varphi =0$, unless stated explicitly otherwise.

The fact that the Carnot cycle realizes the maximum possible efficiency of the heat engine is considered as its remarkable feature.
This theoretical statement provides the crucial criterion in search for the new materials for realization of the effective thermoelectric generator characterized by low heat losses. These losses occur due to the dissipation processes taking place in the working body related to its electrical resistivity and thermal conductivity.

\begin{figure}
\centering
\includegraphics[width=0.8\linewidth]{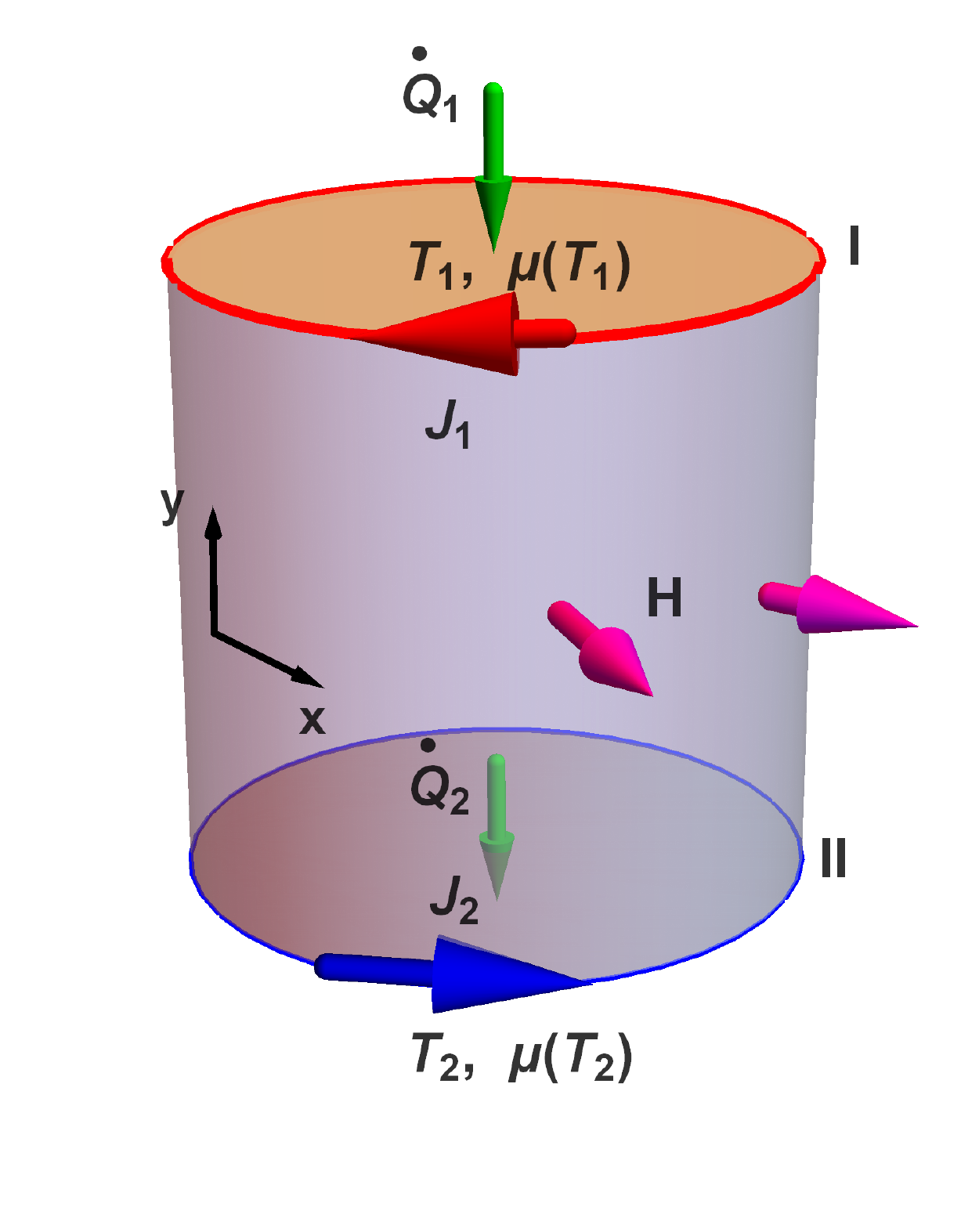}
\caption{The Nernst heat engine in the cylindrical Laughlin's geometry. Electrons are confined on the cylindrical surface in the presence of a magnetic field $H$ applied perpendicularly to the surface. The upper (I) and lower (II) edges are kept at the temperatures $T_1$ and $T_2$. The heat flow is entering through the upper edge  $\dot{Q}_1$ and it leaves the system
via the lower edge  $\dot{Q}_2$ as shown by the green arrows.
The red and blue arrows correspond to the diamagnetic edge currents $J_{1,2}$ that for
$T_1 \neq T_2$ do not compensate each other and result in finite tangential net Nernst current.   }
\label{fig:Laughlin}
\end{figure}

In this article we propose a new type of a {\it gedanken} heat engine based on the Nernst effect realized in the Laughlin geometry \cite{Laughlin1981PRB}
(see Fig.~\ref{fig:Laughlin}).
 We assume that the cylinder surface is an effective insulator, so that no longitudinal charge transport occurs along the cylinder generatrix.
Nevertheless, the edges  remain conducting and
the  non-dissipative  diamagnetic  edge  currents flow \cite{Teller1931}.
The Laughlin geometry allows to visualize these currents making the charged particles flowing along the closed loops in the opposite directions.
Note that the magnetic field direction fully determines the chirality of the edge currents. There are always two counterpropagating
currents at the two edges of the cylinder. One cannot separate them even if cutting the cylinder in two.

Once the edges of the cylinder are put in contact with thermal bathes characterized by the temperatures $T_1$ and $T_2$, the  heat  current   along cylinder generatrix is  induced. We neglect the dissipation of energy during the heat transfer.
The magnitude of the diamagnetic edge currents is temperature dependent, which is why it turns out to be different at the cold and hot edges of the cylinder. For this reason, a total non-zero tangential circular current flows in the system if the temperatures at two edges are different. One can consider the generation of this current as the response of the system to the magnetic field and the temperature difference applied, that constitutes a manifestation of the Nernst effect (see, e.g., Ref.~\cite{Behnia2015book}) in the Laughlin geometry.
One can also consider the model system discussed here as the heat engine where the current is generated due to the difference of chemical potentials $\mu(T_1)$, and $\mu(T_2)$.

\section{Edge currents}

Let us consider the geometry proposed by Laughlin
for the interpretation of the quantum Hall effect \cite{Laughlin1981PRB} that is modified here to address the Nernst effect.
We assume that electrons are confined on a conducting cylindrical surface in the presence of a magnetic field $H$ applied perpendicular to the surface in each of its point
(see Fig.~\ref{fig:Laughlin}). We assume that the edges of the cylinder
are kept at equilibrium with the thermal baths of the temperatures
$T_{1} > T_2$. Under these conditions the working fluid, which consists of an electron gas, is thermalized at both ends at two different temperatures.
	
We are interested here in the regime of classically strong magnetic fields, where the energy separation between
the neighboring Landau levels exceeds their broadening, yet remain
small with respect to the Fermi energy: $ \Gamma  \ll k T \lesssim \hbar\omega_{c} \ll E_{F}$, where $T$ is temperature, $\Gamma$ is the impurity level smearing, $\omega_{c}$ is the cyclotron
frequency, $E_{F}$ is the Fermi energy. In what concerns the requirements
to the cylinder geometry, we assume that its circumference  $L$ and width
$W$ greatly exceed the magnetic length $l_{B}=\sqrt{\hbar c/|e|B}$.
	
As follows from the numerous considerations of the relevant problem
done in the Hall-bar geometry \cite{Teller1931,Obraztsov1965,Hajdu1974ZP,Mineev2007PRB,2020PNAS}
the currents flow along the edges of the conducting layer within the depth of the order of $\widetilde{y}_{0} \sim l_B \sqrt{E_F/(\hbar \omega_c)}$.	
We assume that the temperature gradient is small enough, so that on the scale
of $\widetilde{y}_{0}$ the temperature remains constant.

The value of the edge current can be related to the grand thermodynamical potential $\Omega_L$
per unit area	
\begin{equation}
\label{eq:currentpotent}
\begin{split}
J\left(T,\mu,H\right) =\frac{c}{H }\Omega_{L}\left(T,\mu,H \right),
\end{split}
\end{equation}
where
\begin{equation}
\label{eq:OmegaL}
\begin{split}
& \Omega_{L}\left(T,\mu,H\right)=-2kT\frac{|e|H}{2\pi\hbar c} \\
& \times \sum_{n=0}^{\infty}\ln\left[1+\exp\left(\frac{\mu\left(T\right)-\hbar\omega_{c}\left(n+1/2\right)}{kT}\right)\right].
\end{split}
\end{equation}
Let us stress that in spite of the specifics of the spectrum of the
edge states of electrons the summation in Eq.~(\ref{eq:OmegaL})
can be done  over the spectrum of Landau levels calculated for the infinite system \cite{Obraztsov1965,Hajdu1974ZP,Mineev2007PRB,2020PNAS}.
The spin degeneracy of the electron gas under study is postulated, which results in the  appearance of the factor of 2 in Eq.~(\ref{eq:OmegaL}).
Note that the sign in Eq.~(\ref{eq:currentpotent}) is the matter of convention: In the chosen form it corresponds the direction of the
current flowing along the upper edge of the cylinder.	
	
In the limit of low temperatures $kT\ll\mu\left(T\right)$,
the exponential term in the argument of the logarithm dominates over
1 in Eq.~(\ref{eq:OmegaL}), so that the expression for the current reduces to \cite{2020PNAS}
\begin{equation}
J^{\mathrm{2DEG}}(T,\mu,H)=-\frac{|e|}{\pi\hbar^{2}}\frac{\mu^{2}\left(T\right)}{2\omega_{c}}.
\label{eq:musquare}
\end{equation}

If in the material of the cylinder electrons are characterized by a Dirac spectrum as it happens in graphene, the energy spectrum of Landau levels differs from the equidistant ladder ($E_n = \pm \sqrt{2 n \hbar |e| B v_F^2/c }$), and the summation in $\Omega_{L}$ results in
$J^{\mathrm{gr}}(T,\mu,H)=-(c/H) |\mu (T)|^{3}/(3 \pi \hbar^2 v_F^2),$
where $v_F$ is the Fermi velocity.

If the temperatures of the edges are equal, $T_1 = T_2$,
the currents compensate each other, yet jointly they create a diamagnetic response
to the magnetic field applied.

If
the temperatures of the edges are different,
a full tangential circular current (Nernst current) flows in the cylinder.
Its value is determined by the difference of two edge currents:
\begin{equation}
\label{eq:currdifference}
J_{\mathrm{tot}}=J \left(T_{1}\right)-J\left(T_{2}\right).
\end{equation}
As follows from Eq.~(\ref{eq:musquare}) the direction of the total current
depends on the relative values of the chemical potentials $\mu(T_1)$ and $\mu(T_2)$.
Both of them oscillate as functions of the magnetic field $H$ \cite{Champel2001PM,Lukyanchuk2011PRL}
(for the experimental observation of this effect see Ref.~\cite{Nizhankovskii2011PRes}).
Thus the direction of the total
current may be switched by changing the value of $H$.
In Figs.~\ref{fig:Laughlin} and \ref{fig:gedanken} we assume
that $\mu (T_1) < \mu (T_2)$.

It is striking to note that these oscillations represent a perfect illustration
of the Le~Chatelier-Braun principle, characterized by the coupling effects between
the intensive parameters of the system.
In the present case we observe the direct coupling between two of them,
temperature and chemical potential, which are themselves modified by
the presence of the magnetic field.
It has been shown  that under given conditions these couplings can lead
to spontaneous oscillations \cite{Goupil2016PRE}.
It is sufficient that such oscillations to be governed by only two parameters
for many systems.
Yet, all suggests that increasing the number of these parameters makes the spontaneous
oscillations more likely.

We stress that our argumentation only applies to the ballistic transport regime where
electron-phonon scattering is negligible. In a high temperature limit, the scattering
would eventually suppress dissipationless edge currents.
A hotter side would be characterized by a lower current magnitude in this case.

Let us also note that the numerical simulations of the Nersnt effect
in a fluctuating superconductor in the Laughlin geometry \cite{Sarkar2017NJP}
shows indeed that the total circular current appears when the edges are kept at the different
temperatures.

As it was mentioned above, the edge currents are formed by the skipping electrons located
within the stripes of the width $\widetilde{y}_{0}$ near the edges, while the electrons at the surface of the cylinder far from the edges are confined to their cyclotron orbits and do not contribute to the electric current \cite{Obraztsov1965,Hajdu1974ZP,Mineev2007PRB,2020PNAS}.
The positions of the centres of cyclotron orbits  at the generatrix are determined by the momentum quantum number $p_x$, and they are fixed.
Due to the  non-zero temperature $k T \lesssim  \hbar \omega_c$ several quantum states with the Landau levels close to the Fermi level may be available for the electron transport. Electron-electron or electron-phonon interaction can result in the transitions of the electron, rotating around the same center, between the Landau levels. As a result, the heat from the hot edge to the cold one can flow  by means of subsequent resonant transitions of the electrons from higher to lower Landau levels without their macroscopic displacements along the generatrix.
In this way, the heat transfer  without the ballistic charge transfer occurs.
We also point out that the non-trivial topology of the Laughlin cylinder results in the highly unusual diamagnetic response: The non-equal circular currents along the edges induce a magnetic field parallel to the axis of the cylinder and perpendicular to the external magnetic field. We admit that the considered geometry of the external magnetic field implies the existence of magnetic monopoles and cannot be realized in practice, strictly speaking.
	
\section{Thermodynamic treatment of the Nernst effect}

\begin{figure}
\centering
\includegraphics[width=.9\linewidth]{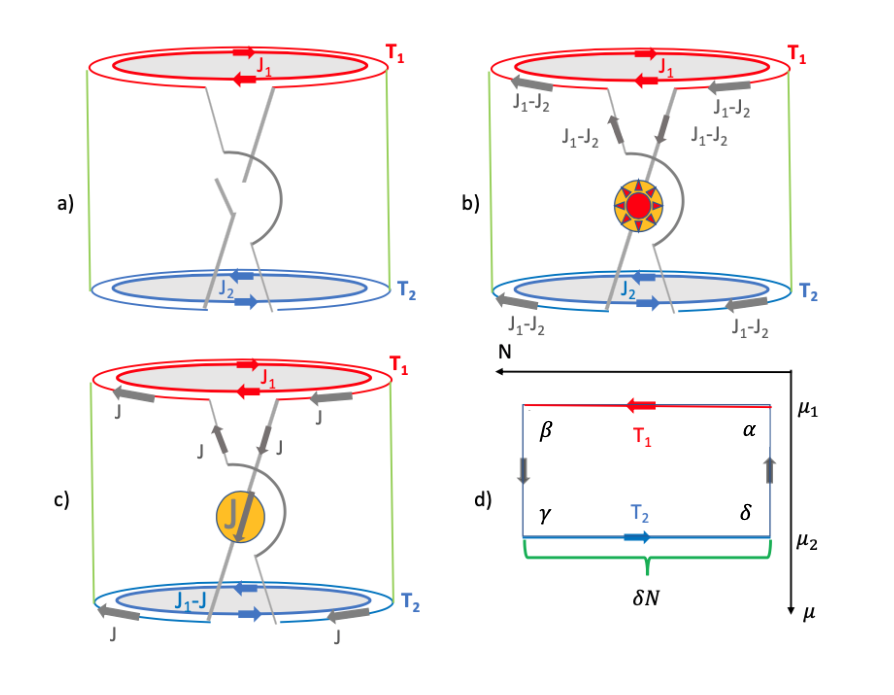}
\caption{ A scheme of the {\it gedanken } Nernst heat engine based on a Laughlin cylinder whose edges are immersed in thermal baths of different temperatures. Panel (a) shows the broken circuit regime, were no work is produced, panel (b) shows the active regime, where the work is produced
by a current flowing through the bulb. This current is given by the difference of two edge currents. It is compensated by a thermal current flowing through the cylinder in the opposite direction.
Panel(c) shows the Ettingshausen heater based on the inverse effect: The current produced by the generator introduces the imbalance of edge currents that is compensated due to the induced temperature imbalance between the two edges. If the lower edge is kept at a constant temperature of a bath, the temperature at the upper edge increases. Panel(d) shows the $\mu$ -- $N$ diagram for the Nernst heat engine cycle.\\
${\pmb \alpha} {\bf \to} \, {\pmb \beta}$ (isothermal process):the entering heat flow $Q_1$  from the higher edge kept at the constant temperature $T_1$ to the  electronic system at the upper  edge results in the increase of the kinetic energy of the electrons at  the upper edge and the increase of their entropy, so that
$Q_{1} =T_1 S_1=  \mu_{1}(- \delta N)$.  \\
${\pmb{\beta} {\bf \to} \, {\pmb \gamma}}$: the work is produced while electrons pass
through the load situated between the edges.
As the process is adiabatic and reversible, the incoming and outgoing
entropy rates $\dot{S}_1 = \dot{S}_2 = \dot{S}$ [see Eq.~(\ref{eq:entropy-rates})]. \\
${\pmb  \gamma} {\bf \to} \, {\pmb \delta}$ (isothermal process): the cooler absorbs the heat flow coming from the lower edge:
$Q_{2} =  T_2 S_2  = \mu_2 (- \delta N)$. \\
${\pmb \delta} {\bf \to} \, {\pmb \alpha }$: corresponds to the adiabatic return to the initial state.
The entropy flow $ - \dot{S}_1 = - \dot{S}_2 = \dot{S}$.}
\label{fig:gedanken}
\end{figure}

The  diamagnetic tangential circular edge currents $J\left(T_{1}\right)$ and $J\left(T_{2}\right)$  flow along the upper and lower loops,
respectively [see Eqs.~(\ref{eq:currdifference}) and (\ref{eq:currentpotent})] (see Fig.~\ref{fig:Laughlin}). As discussed above,
no ballistic longitudinal charge transport occurs along the cylinder generatrix, while the thermal transport due to the resonant tunneling of electrons from the  upper Landau levels in the bottom part of the cylinder to the lower Landau levels in the top part of the cylinder is possible in the presence of a gradient of the electrostatic potential along the generatrix of the cylinder.
To extract the  work done by the system
let us set up the {\it gedanken experiment} as it is shown in Fig.~\ref{fig:gedanken}~(a)
(cf. the spiral thermobattery  \cite{Anatychuk2007JT}).

The extremes of the cylinder are immersed in thermal baths kept at the temperatures $T_1$ and $T_2$. The heat budget at the boundaries of the system is provided by  the  heat flow   $\dot{Q}_1$
entering through upper edge and outgoing through the lower edge $\dot{Q}_2$.
The considered heat engine is supposed to be fully reversible, which requires the heat  transfer along the cylinder generatrix to be non-dissipative. This implies that  the losses occurring due to electrical and thermal conductivities are neglected.

There is a profound analogy between the classic heat engine that uses
a molecular gas as the working body and a thermoelectric generator,
which employs the degenerated Fermi gas instead Refs.~\cite{Samoylovich2007book,Goupil2015book}.
Following the same logic and based on the results presented in the previous sections we will show how the ideal Nernst heat
engine may work.
The role of the working body in this construction also plays a degenerated Fermi gas localized at two extremes of the cylinder.
The  different temperatures
of the edges result in slightly different values of the chemical potentials  $\mu_1 =\mu(T_1)$ and $\mu_2 = \mu(T_2)$ (contrary to strongly different electrochemical potentials of two dissimilar metals of a thermoelectric couple)
that leads to the entropy transfer.

Although in a thermoelectric  device there is no periodical mechanical motion of the working body, it is instructive to represent its operation process in terms of a ``thermoelectric cycle'' in the $\mu$ -- $N$ diagram \cite{Vining1997}, with $N$ being the number of particles.
A similar approach was used in Ref.~\cite{Chua2002PRE}, where the thermoelectric chiller and generator cycles were
considered using the
temperature -- entropy flux diagrams. Recently, in
Ref.~\cite{Alicki2017AP},  the photovoltaic conversion process
in a solar cell was also represented as a thermodynamic cycle.
This cycle leads to  maintaining the electrochemical potential difference, which would otherwise be reduced to zero because of the recombination. The constant supply of energy through the device leads to the maintenance of this potential difference. Hence,
for the description of our {\it gedanken experiment} schematically illustrated in
Figs.~\ref{fig:gedanken}~(a) and \ref{fig:gedanken}~(b), we employ the language of a
``thermoelectric cycle'' and present the corresponding $\mu$ -- $N$
plot in Fig.~\ref{fig:gedanken}~(d).

In the case of a totally reversible Carnot cycle, the entering and outgoing
entropy rates related to the heat budget are equal
\begin{equation}
\label{eq:entropy-rates}
\dot{S} = \dot{S}_{1,2} = \frac{\dot{Q_1}}{T_1}=  \frac{\dot{Q}_2}{T_2}.
\end{equation}
The ``transported'' entropies  $S_{1,2}$ are nothing else but the entropies of the working substances,
i.e. the electrons at the edges.
The heat budget difference then is given by
\begin{equation}
\Delta \dot{Q}=\dot{Q}_1-\dot{Q}_2= \dot{S}(T_1-T_2) = \eta_C \dot{Q}_1 ,
\label{heatprod}
\end{equation}
where $\eta_C = 1 - T_2/T_1$ is the Carnot efficiency.

The work of a traditional heat engine is determined by the area restricted by the closed working cycle in the phase space:
pressure ($P$) and volume ($V$).
Following the arguments developed in Ref.~\cite{Goupil2015book} one can replace $P \to \mu$ and $V \to N$ and express the work produced during one cycle as follows:
\begin{equation}
\label{work}
W  = (\mu_2  - \mu_1) \delta N.
\end{equation}
Here $\delta N$ is the number of particles that crosses
the load during one cycle, i.e.,
\begin{equation}
\label{period}
\delta N = -\frac{J_1- J_2}{|e|} \tau =-\frac{J_{\mathrm{tot}}}{|e|}\tau,
\end{equation}
where $\tau$ is the period of  the cycle. Note that
Eqs.~(\ref{work}) and (\ref{period}) can be easily reduced to the textbook formula for the work of electric current represented as a product of current, voltage and time keeping in mind that the difference of chemical potentials between two edges of the cylinder is exactly compensated by the difference of their electrostatic potentials, in the stationary regime.

One finds from the energy conservation law $ {W} = \Delta Q$
and Eq.~(\ref{heatprod}) that
\begin{equation}
S (T_1 - T_2) = (\mu_2 - \mu_1) \delta N .
\label{eq:entr}
\end{equation}
Here $S$ is the budget of the total entropy that
flows through the system during  one cycle.
We stress that $S$ is the entropy flow through the system in contrast to
{\it the entropy of the system}, which is a state function that comes to its initial value after every cycle.

The total persistent tangential current  (\ref{eq:currdifference})
in the case of a 
two-dimensional electron gas
[see Eq.~(\ref{eq:musquare})] can be written in the form
\begin{equation}
\label{eq:currdif}
J_{\mathrm{tot}}= - \frac{|e|}{\pi\hbar^{2}}\frac{\mu_1^{2}-\mu_2^{2}}{2\omega_{c}}=\frac{nc}{H}(\mu_{2}-\mu_{1}),
\end{equation}
where we identified $(\mu_{1} + \mu_{2})/2 \approx \pi n\hbar^{2}/m$
and $n$ is the carriers concentration for a 2D system.

Comparing Eqs.~(\ref{eq:entr}) and (\ref{eq:currdif}) one finds that the expression for the total current can be presented in a rather simple and  universal form:
\begin{equation}
\label{eq:currdeltauni}
J_{\mathrm{tot}}=\frac{c  n}{H} \left( \frac{S}{\delta N } \right) (T_{1}-T_{2}) =
\frac{c \mathcal{S}}{H} (T_1 -T_2).
\end{equation}
Here $S/\delta N$ is the entropy budget per carrier and
$\mathcal{S} = (S/\delta N) n$ is the entropy budget per unit area.
Note that the ratio  $\mathcal{S}/H$ in front of the temperature difference in the second equality of
Eq.~(\ref{eq:currdeltauni}) is nothing else but the ratio of the  full entropy budget per magnetic flux penetrating the area. This parameter seems to be as fundamental as the conventional filling factor \cite{Hajdu1994book}.

For the specific case of graphene characterized by a linear (Dirac) energy spectrum of electrons,
the relationship between carrier density (imbalance)
and chemical potential reads as
$n= \mu^2 \mbox{sgn} (\mu) /(\pi \hbar^2 v_F^2 ) $.
One can see that	the above derivation of  Eq.~(\ref{eq:currdeltauni})  remains valid.
	
The proposed scheme of a {\it gedanken experiment} allows to directly measure both the diamagnetic currents and the Nernst coefficient.

\section{Reverse cycle: the Ettingshausen effect}

Another type of instructive {\it gedanken experiment} can be proposed in the Laughlin geometry. In this case, an ideal current generator replaces the load [see Fig. \ref{fig:gedanken}(c)]. Only the lower edge of the cylinder is kept in a bath of a constant temperature,
while the temperature of the upper edge can vary.  Once
the circuit is closed and the ideal current generator is switched on, the edge currents flowing along the upper and lower edges start getting imbalanced. Indeed, if an ideal current generator in the intermediate chain feeds the circuit by the current $J_1-J_2$, the current $J_2$ is fixed in the lower circuit. This last circuit stays at the same magnetic field as the upper one, which is why the given value of the flowing edge current requires the equilibrium temperature $T_2 < T_1$ (i.e., $\mu_2 > \mu_1$). Hence, the entire system operates as a heater. One can easily convert it to a refrigerator by inverting the direction of the generated current.
The work of the current generator is spent to pump heat from the thermal bath to the upper edge of the cylinder.

During one cycle the fraction of heat pumped to
the bottom reservoir is $Q_2 = T_2 (-S)$, where the entropy budget
is the same as above up to the sign. Thus the heat
flow $\dot{Q}_2 = Q_2/\tau $ supplied to the bottom reservoir reads
\begin{equation}
\label{heat-flow}
\dot{Q}_2  = T_2 \frac{c n}{H} \frac{S}{\delta N} \frac{\mu_2 - \mu_1}{|e|} =  -T_2 \frac{c \mathcal{S}}{(-H)} \left(\frac{\mu_2 - \mu_1}{e} \right),
\end{equation}
where we used  Eqs.~(\ref{period})  and  (\ref{eq:currdif}) relating the cycle period and the total current.
As one can observe, the total electric current (\ref{eq:currdeltauni})
and heat current (\ref{heat-flow}) are linked to each other by the Onsager relation \cite{Callen1948PR}:
\begin{equation}
\label{Onsager}
\frac{J_{\mathrm{tot}}}{T_1 - T_2} = - \frac{1}{T_2}\frac{\dot{Q}_2}{(\mu_1 - \mu_2)/e} =    \frac{c \mathcal{S}}{H}.
\end{equation}

\section{Thermodynamic versus microscopic approaches}	

In the seminal paper \cite{Obraztsov1965}
Obraztsov obtained the version of Eq.~ (\ref{eq:currdeltauni}).
He studied a Hall-bar that corresponds
to the cut of the Laughlin cylinder in Fig.~\ref{fig:Laughlin} along the
generatrix.
The consideration in Ref.~\cite{Obraztsov1965} is based
on the requirements of thermal equilibrium and electroneutrality
of the charged system, i.e., the constancy of the electrochemical potential
$e \varphi (\mathbf{r}) + \mu (\mathbf{r}) = \mbox{const.}$
and temperature $T (\mathbf{r}) = \mbox{const}$.

The electric current density in this approach appears as a response
to the perturbation
\begin{equation}
\label{response}
j_x = \sigma_{xy} \left(E_y + \frac{1}{e} \nabla_y \mu \right) - \beta_{xy} \nabla_y T.
\end{equation}
Here $E_y = - 1/e \nabla_x \varphi$, $\sigma_{xy}$ is the Hall conductivity,
and $\beta_{xy}$ is the off-diagonal part of the thermoelectric tensor \cite{Samoylovich2007book}.
Assuming that the temperatures of the two edges are close  to each other, $T_2 = T_1 + \Delta T$,
and that the height of the sample is $l_y$ Obraztsov obtained
from Eq.~(\ref{eq:currentpotent}) for the
microscopic density of electrical current
\begin{equation}
\label{current-smeared}
j_x = \frac{J(T+\Delta T) - J(T)}{l_y} = \frac{c}{H} \left(\frac{d \Omega_L}{d T}\right) \nabla_y T.
\end{equation}
Let us note that at this point the author simplified the model
assuming that the currents are not localized at the edges but, in contrast, are distributed over the entire sample.

Substituting Eq.~(\ref{current-smeared}) in
Eq.~(\ref{response}) and using the condition $E_y =0$ one finds
\begin{equation}
\label{beta-Obraztsov}
\beta_{xy} = \frac{\sigma_{xy}}{e} \frac{d \mu}{d T} - \frac{c}{H} \frac{d \Omega_L}{d T}.
\end{equation}
Since the differential of the thermodynamic potential per unit area
$
d \Omega_L (T, \mu,H) = - \mathcal{S} d T - n d \mu - \mathcal{M} d H,
$
where $\mathcal{S}$ and $\mathcal{M}$ are the entropy and magnetization per unit area,
one obtains
$
d \Omega_L /d T = - n  d \mu /d T - \mathcal{S}.
$
From here, using the classical expression for the Hall conductivity, $\sigma_{xy} = - c e n/H$ the author arrived
to the final result \cite{Obraztsov1965,Hajdu1974ZP,Oji1985PRB}
\begin{equation}
\label{Obraztsov-NE}
\beta_{xy} = \frac{c \mathcal{S}}{H},
\end{equation}
which connects the current density with the temperature gradient, entropy per unit square and magnetic field. Comparing it with Eq.~(\ref{eq:currdeltauni}) one can see that both equations are essentially similar, yet deriving Eq.~(\ref{eq:currdeltauni})  we did not use any assumptions regarding the nature of the material (e.g. such as an explicit formula for the classical Hall conductivity).
Due to the reversibility of the cycle the entropy of the system remains constant and it coincides with the entropy
budget introduced above.

\section{Conclusions}

We discussed the model of an ideal reversible thermodynamic cycle for the visualization of the Nernst and Ettingshausen effects in the Laughlin geometry. This geometry allows for eliminating the kinetic contribution to the Nernst effect.
We express the off-diagonal thermoelectric coefficients  through the ratio of the entropy budget per magnetic flux.

While the realization of the proposed {\it gedanken experiment} in practice is challenging in the Laughlin cylinder geometry, a similar phenomenology of the edge currents may be studied in a Corbino disk whose inner and outer edges are immersed in thermal baths of different temperatures as discussed in Ref.~\cite{2020PNAS}.
In particular, the Nernst effect in the electron gas
with a parabolic carrier dispersion and in graphene was analyzed in that work, with
the main focus on the experimentally observable effect, such as oscillations of the magnetization.
Yet, the analysis of the {\it gedanken experiment} discussed in the present work would become more
cumbersome due to the mismatch in the radii of the two edges.

Our approach enlightens the profound thermodynamic origin of the relation between the Nernst effect and magnetization currents, obtained in Ref.~\cite{Obraztsov1965} in the framework of a microscopic model for a specific fermionic system.

Let us stress, that the model behind  Eq.~(\ref{beta-Obraztsov}) (see also  Eq.~(26) in Ref.~\cite{Obraztsov1965})
offers some significant advantages over the frequently employed Kubo approach
\cite{Anselm1961,Obraztsov1964,Obraztsov1965,Smrcka1977JPC,Jonson1984PRB,
Oji1985PRB,Ruzin1997PRB,SSVG2009PRL,Fink2009,Reizer2011PRL,Qin2011PRL}.
Indeed, the off-diagonal components of the thermoelectric tensor
calculated in the kinetic approach and expressed in terms of the correlator of electric current --
heat current operators are incomplete, which is why the calculations involving these components often suffer from fundamental contradictions.
For example, the Nernst coefficient of a non-interacting electron system subjected to the
quantizing magnetic field calculated in the Kubo approach does not satisfy the Onsager principle of the symmetry of kinetic coefficients Ref.~\cite{Anselm1961}. The corresponding diagrammatic calculations of the Nernst signal induced by fluctuation Cooper pairs (see, e.g., \cite{Sarkar2016AP})
close to the second critical field and low temperatures demonstrates \cite{SSVG2009PRL} the unphysical divergence contradicting the third law of thermodynamics. The same divergence appears in the temperature dependence of the
spin Nernst effect in the vicinity of  zero temperature \cite{GSV2014}.
A convenient way to resolve these contradictions was indicated by Obraztsov half a century ago in his seminal paper \cite{Obraztsov1964}. A self-consistent and physically correct description of a variety of thermomagnetic phenomena is achieved accounting for the magnetization currents in addition to the kinetic  contributions following from the Kubo approach. Indeed, the proposed thermodynamic approach based on Eqs.~(\ref{eq:currdeltauni}) and (\ref{beta-Obraztsov})  is free of the above mentioned internal inconsistencies.

We do believe that the {\it gedanken experiments} considered here will help to put a final point in the long discussion \cite{Anselm1961,Obraztsov1964,Obraztsov1965,Smrcka1977JPC,Jonson1984PRB,
Oji1985PRB,Ruzin1997PRB,SSVG2009PRL,Fink2009,Reizer2011PRL,Qin2011PRL}  on the role of magnetization currents in the thermomagnetic phenomena.

\begin{acknowledgments}

A.V.K., S.G.Sh., and A.A.V. are grateful to B.L.~Altshuler
for the attraction of our attention to the special role of
Laughlin geometry in consideration of the Nernst effect.
S.G.Sh. thanks V.M.~Loktev and A.A.~Snarskii for useful discussions. A.A.V.
is grateful to Yu.M.~Galperin and V.B.~Shikin for illuminating discussions.
A.A.V. and S.G.Sh. acknowledge the hospitality of Westlake University, where this work was started.	
A.A.V. acknowledges support by  the European Union's Horizon 2020  research and innovation program  under the
Grant Agreement No.~731976 (MAGENTA). 	S.G.Sh. acknowledges a support by the National Research Foundation of Ukraine grant  (2020.02/0051) ``Topological phases of matter and excitations in Dirac materials, Josephson junctions and magnets''.
A.V.K. acknowledges the Project No.~041020100118 and Program 2018R01002
supported by Leading Innovative and Entrepreneur Team Introduction Program of Zhejiang.

\end{acknowledgments}

\end{document}